\documentclass[12pt,thmsa,notitlepage]{article}
\usepackage{amsmath}
\usepackage{amssymb}
\usepackage{amsfonts}
\usepackage{sw20lart}
\usepackage[onehalfspacing]{setspace}

\input{tcilatex}

\begin{document}

\title{${\LARGE \Delta Q=2}$ Hyperon Decays}
\author{{\large Ling-Fong Li} \\
{\small Department of Physics, Carnegie Mellon University, Pittsburgh, PA
15213}}
\maketitle

\begin{abstract}
The hyperon decays with 2 negative charged leptons in the final states can
be of interest as possible tests for the Majorana netrinos. The study of
these decays are complementary to the double $\beta $ decays of heavy nuclei.
\end{abstract}

It is now established that neutrinos do have non-zero masses\cite{Kamiokande}%
\cite{KamLAND}\cite{SNO}. The next step is to study the properties of these
massive neutrinos. One interesting question is whether they are Majorana or
Dirac neutrinos. Over the years the search for neutinoless double $\beta $
decays of some heavy nulcei\cite{double beta} is to test the possibility
that neutrinos are Majorana \ particles. However, theoretical analysis for
these processes is hampered by the complicate nuclear matrix elements which
are very difficult to calculate reliably. As an alternative, we can study
leptonic decays of the hyperons, $B_{1}\rightarrow B_{2}ll^{\prime }$. Here $%
l$ and $l^{\prime }$ are same sign leptons, like $e^{-}e^{-},\mu ^{-}\mu
^{-},$ or $e^{-}\mu ^{-}.$ The study of these processes can serve as
completement to the nuclear double $\beta $ decays. Some upper bounds
already exist for some of these processes\cite{Littenberg}\cite{CPhyperon}.

We can classify the possible hyperon decays by the change in the strangeness.

\begin{enumerate}
\item $\Delta S=0$%
\begin{equation*}
\Sigma ^{-}\rightarrow \Sigma ^{+}e^{-}e^{-},\qquad M\left( \Sigma
^{-}\right) -M\left( \Sigma ^{+}\right) =8MeV
\end{equation*}

\item $\Delta S=1$%
\begin{equation*}
\Sigma ^{-}\rightarrow pe^{-}e^{-},\qquad M\left( \Sigma ^{-}\right)
-M\left( p\right) =257MeV\text{ \ \ \ \ \ \ \ \ \ \ \ \ }
\end{equation*}%
\begin{equation*}
\Sigma ^{-}\rightarrow p\mu ^{-}e^{-},\qquad M\left( \Sigma ^{-}\right)
-M\left( p\right) -M\left( \mu \right) =151MeV
\end{equation*}%
\begin{equation*}
\Sigma ^{-}\rightarrow p\mu ^{-}\mu ^{-},\qquad M\left( \Sigma ^{-}\right)
-M\left( p\right) -2M\left( \mu \right) =46MeV
\end{equation*}%
\begin{equation*}
\Xi ^{-}\rightarrow \Sigma ^{+}e^{-}e^{-},\qquad M\left( \Xi ^{-}\right)
-M\left( \Sigma ^{+}\right) =132Mev
\end{equation*}%
\begin{equation*}
\Xi ^{-}\rightarrow \Sigma ^{+}\mu ^{-}e^{-},\qquad M\left( \Xi ^{-}\right)
-M\left( \Sigma ^{+}\right) -M\left( \mu \right) =26Mev
\end{equation*}
\end{enumerate}

$\Delta S=2$%
\begin{equation*}
\Xi ^{-}\rightarrow pe^{-}e^{-},\qquad M\left( \Xi ^{-}\right) -M\left(
p\right) =383MeV\text{ \ \ \ \ \ \ \ \ \ \ \ \ }
\end{equation*}%
\begin{equation*}
\Xi ^{-}\rightarrow p\mu ^{-}e^{-},\qquad M\left( \Xi ^{-}\right) -M\left(
p\right) -M\left( \mu \right) =277MeV
\end{equation*}%
\begin{equation*}
\Xi ^{-}\rightarrow p\mu ^{-}\mu ^{-},\qquad M\left( \Xi ^{-}\right)
-M\left( p\right) -2M\left( \mu \right) =172MeV
\end{equation*}%
\begin{equation*}
\Omega ^{-}\rightarrow \Sigma ^{+}e^{-}e^{-},\qquad M\left( \Omega
^{-}\right) -M\left( \Sigma ^{+}\right) =483MeV\text{ \ \ \ \ \ \ \ \ \ \ \
\ }
\end{equation*}%
\begin{equation*}
\Omega ^{-}\rightarrow \Sigma ^{+}\mu ^{-}e^{-},\qquad M\left( \Omega
^{-}\right) -M\left( \Sigma ^{+}\right) -M\left( \mu \right) =377MeV
\end{equation*}%
\begin{equation*}
\Omega ^{-}\rightarrow \Sigma ^{+}\mu ^{-}\mu ^{-},\qquad M\left( \Omega
^{-}\right) -M\left( \Sigma ^{+}\right) -2M\left( \mu \right) =272MeV
\end{equation*}%
\qquad \qquad

We see that the $\Delta S=0$ decay $\Sigma ^{-}\rightarrow \Sigma
^{+}e^{-}e^{-}$ has very small phase space to be of interest and there are
many other processes which have comparable phase space as the usual hyperon
leptonic decays.

\underline{\textbf{Effective interaction}}s

The Majorana neutrinos can give rise to the decay processes discussed here
just like the case in the nuclear double $\beta $ decays, e.g.\FRAME{dtbpF}{%
3.5898in}{2.1525in}{0pt}{}{}{figm0.jpg}{\special{language "Scientific
Word";type "GRAPHIC";maintain-aspect-ratio TRUE;display "USEDEF";valid_file
"F";width 3.5898in;height 2.1525in;depth 0pt;original-width
1.7608in;original-height 1.0456in;cropleft "0";croptop "1";cropright
"1";cropbottom "0";filename 'Figm0.jpg';file-properties "XNPEU";}}But there
can also be some new interaction where a new doubly charged scalar particle
couples to dileptons as shown in the figure\cite{Cheng-Li},\FRAME{dtbpF}{%
3.6979in}{2.3134in}{0pt}{}{}{figm1.jpg}{\special{language "Scientific
Word";type "GRAPHIC";maintain-aspect-ratio TRUE;display "USEDEF";valid_file
"F";width 3.6979in;height 2.3134in;depth 0pt;original-width
2.0548in;original-height 1.1355in;cropleft "0";croptop "1";cropright
"0.8916";cropbottom "0";filename 'figm1.jpg';file-properties "XNPEU";}}Of
course there are other mechanism for these decays and they will undoubtly
involve new particles and/or new interactions. Thus the study of these
processes can shed some light on either Majorana nature of the neutrinos or
uncover some new physics beyond the standard model. Here we will just
discuss some general features of these processes without going into details
of each possible models. It is reasonable to assume the effective
interaction for the processes of intererst can be written as%
\begin{eqnarray*}
L_{\beta \beta } &=&\dfrac{G_{F}^{2}}{\Lambda _{\beta \beta }}\left[ c_{1}(%
\overset{\_}{u}\Gamma _{i}d)(\overset{\_}{u}\Gamma _{j}d)+c_{2}(\overset{\_}{%
u}\Gamma _{i}d)(\overset{\_}{u}\Gamma _{j}s)+c_{3}(\overset{\_}{u}\Gamma
_{i}s)(\overset{\_}{u}\Gamma _{j}s)\right] \\
&&\times \left[ d_{1}(\overset{\_}{e}\Gamma _{k}e^{c})+d_{2}(\overset{\_}{e}%
\Gamma _{k}e^{c})+d_{3}(\overset{\_}{e}\Gamma _{k}\mu ^{c}+\overset{\_}{\mu }%
\Gamma _{k}e^{c})\right]
\end{eqnarray*}%
Here $\Lambda _{\beta \beta }$ is a parameter which charaterizes the physics
scale responsible for these processes and $c_{i},d_{i}$ are dimensionless \
parameters which represent different the interaction strengths for various
channels. These parameters depend on specific model and certainly will
contain some unknown parameters. Also $\Gamma _{i}^{\prime }s$ are some
combinations of Dirac gamma matrices which depend on the physical mechanism
involved.

Since the matrix elements of the 4-quark operators are difficult to
calculate, we will just get some order of magnitude estimate of these
decays. The hadronic matrix elements of 4-quark operator will be compared to
that of the hyperon leptonic decays,%
\begin{equation*}
R_{h}=\dfrac{\left\langle B_{2}|(\overset{\_}{u}\Gamma _{i}d)(\overset{\_}{u}%
\Gamma _{j}s)|B_{1}\right\rangle }{\left\langle B_{2}|((\overset{\_}{u}%
\Gamma s)|B_{1}\right\rangle }
\end{equation*}%
This ratio has mass dimension 3 and we will write it as%
\begin{equation*}
R_{h}=M^{3}
\end{equation*}%
where $M$ is some mass parameter. It is possible that $M$ is of order of
hadron mass of $1$ $Gev$ or so. We can write the decay rates as%
\begin{equation*}
\dfrac{\Gamma \left( B_{1}\rightarrow B_{2}ll^{\prime }\right) }{\Gamma
\left( B_{1}\rightarrow B_{2}e\nu \right) }\sim \left( \dfrac{G_{F}}{\Lambda
_{\beta \beta }}\right) ^{2}M^{6}\rho 
\end{equation*}%
where $\rho $ represents the phase spaces ratio. If we take $M$ to be of
order of $1$ Gev or so, this ratio is of order of%
\begin{equation*}
\dfrac{\Gamma \left( B_{1}\rightarrow B_{2}ll^{\prime }\right) }{\Gamma
\left( B_{1}\rightarrow B_{2}e\nu \right) }\sim \left( \dfrac{M}{\Lambda
_{\beta \beta }}\right) ^{2}\rho \times 10^{-10}
\end{equation*}%
In the usual hyperon leptonic decays, when the lepton masses are neglected,
the phase space is proportional to 
\begin{equation*}
\Delta ^{5}=\left( M_{1}-M_{2}\right) ^{5}
\end{equation*}%
where $M_{1},M_{2}$ are masses of $B_{1},B_{2}$ respectively. This should
apply also to decay channels with 2 electrons in the final states. But for
cases where there are muons in the final states $\Delta $ should be replaced
by%
\begin{equation*}
q=\sqrt{\left( M_{1}-M_{2}\right) ^{2}-m^{2}}\text{ \ \ \ or \ \ }\sqrt{%
\left( M_{1}-M_{2}\right) ^{2}-4m^{2}}
\end{equation*}%
where $m$ is the muon mass. In most cases, the phase space does not give
large suppression or enchancement except for the decay $\Sigma
^{-}\rightarrow \Sigma ^{+}e^{-}e^{-}$ where the phase space suppression is
of order of $10^{-8}$ or so. In some cases with $\Delta S=2$, there is a
small enchancement due to the fact the mass differences can be larger than
those with $\Delta S=1.$ For example, in the decay $\Xi ^{-}\rightarrow
pe^{-}e^{-}$ the phase space is larger by a factor of $200$ over that of $\
\Xi ^{-}\rightarrow \Sigma ^{0}e^{-}\nu _{e}.$

For the case of Majorana neutrinos, the factor $\dfrac{1}{\Lambda _{\beta
\beta }}$ should contain a factor of Majorana mass of neutrinos $m_{maj}$,%
\begin{equation*}
\dfrac{1}{\Lambda _{\beta \beta }}=\left( \dfrac{m_{maj}}{\Lambda ^{\prime }}%
\right) \dfrac{1}{\Lambda ^{\prime }}
\end{equation*}%
and will probably give a large suppression factor. Thus branching ratio of
order of $10^{-10}$ is probably most optimistic upper bound for these type
of decays and most likely they are much smaller than this.

\end{document}